\documentclass[12pt]{article}

\usepackage{fancyhdr}
\usepackage{epsfig}
\usepackage{cite}
\usepackage{graphicx}
\usepackage[centertags]{amsmath}
\usepackage{theorem}

\usepackage{graphicx}
\usepackage{amssymb}
\usepackage{latexsym}
\usepackage{epic}
\usepackage{pstricks}
\usepackage{color}
\usepackage{mathrsfs}
\usepackage{latexsym}
\usepackage{epic}
\usepackage{braket}
\usepackage{slashed}
\newcommand{\be}{\begin{equation}}
\newcommand{\ee}{\end{equation}}
\newcommand{\bea}{\begin{eqnarray}}

\newcommand{\eea}{\end{eqnarray}}

\numberwithin{equation}{section}
\def\hybrid{\topmargin 22pt    \oddsidemargin 0pt 
      \headheight 0pt \headsep 0pt
      \textwidth 6.5in        
       \textheight 9in         
      \marginparwidth .875in
      \parskip 5pt plus 1pt   \jot = 1.5ex}

\hybrid

\usepackage{caption} 
\newcommand{\alphad}{\dot{\alpha}}
\newcommand{\betad}{\dot{\beta}}

\newcommand{\refe}[1]{Eqn.~(\ref{#1})}

\renewcommand{\thefootnote}{\fnsymbol{footnote}}

\begin{document}
\center{
\begin{flushright} 
DESY 12-006 \\
QMUL-PH-11-21
\end{flushright}
\vspace{1cm}
\begin{center}
{\Large\textbf{Super-Higgs in Superspace}}
\vspace{1cm}

\textbf{Moritz McGarrie$^{1,}$\footnote[2]{\texttt{moritz.mcgarrie@desy.de}} and Gianni Tallarita$^{2,}$ \footnote[3]{\texttt{g.tallarita@qmul.ac.uk}} }\\
\end{center}

{\it{ ${}^1$ 
Deutsches Elektronen-Synchrotron,\\ DESY, Notkestrasse 85, 22607 Hamburg, Germany
}\\
\vspace{0.5cm}
\it{ ${}^2$ Queen Mary University of London\\ Centre for Research in String Theory \\ School of Physics and Astronomy \\ Mile End Road, London, E1 4NS, UK}}

\setcounter{footnote}{0}
\renewcommand{\thefootnote}{\arabic{footnote}}

\abstract{We determine the effective gravitational couplings in superspace whose components reproduce the supergravity Higgs effect for the constrained Goldstino multiplet. It reproduces the known Gravitino sector whilst constraining the off-shell completion.  We show that these components arise by computing the effective action.  This may be useful for phenomenological studies and model building: We give an example of its application to multiple Goldstini.}


\section{Introduction}
The spontaneous breakdown of gobal supersymmetry generates a massless Goldstino \cite{Salam:1974zb}, which is well described by the Akulov-Volkov (A-V) effective action \cite{Volkov:1973ix}. When supersymmetry is made local,  the Gravitino ``eats" the Goldstino of the A-V action to become massive:  The super-Higgs mechanism \cite{Deser:1977uq}.

In terms of superfields, the constrained Goldstino multiplet $X_{NL}$   \cite{Komargodski:2009rz}, is equivalent to the A-V formulation (see also \cite{Kuzenko:2010ni,Kuzenko:2010ef,Kuzenko:2011tj,Dudas:2011kt,Antoniadis:2011xi}).  It is therefore natural to extend the description of supergravity with this multiplet, in superspace, to one that can reproduce the super-Higgs mechanism.     In this paper we write down the most minimal set of \emph{new} terms in superspace that incorporate both supergravity and the Goldstino multiplet in order to reproduce the super-Higgs mechanism of \cite{Deser:1977uq,Rychkov:2007uq} at lowest order in $M_{Pl}$. 

In writing down an effective action we should expect that these terms are generated radiatively.  To check this we explicitly compute the effective action in components and demonstrate that the necessary terms do indeed appear. What is perhaps surprising is that these terms arise from an interplay of the gravitational cosmological constant that appears in the scalar component of the Goldstino multiplet.  We then set the gravitational cosmological constant to cancel the F-term of the Goldstino, a prerequisitive for the super-Higgs mechanism.  In retrospect this should not be so surprising:  in supersymmetry all coefficients should be promoted to background superfields and this includes the cosmological constant.  It it there sensible to suspect that the cosmological constant is contained in a chiral superfield.   In this paper we show that the cosmological constant may naturally be contained within the scalar component of $X_{NL}$.  Furthermore we show that all the soft terms associated with the super-Higgs mechanism may be naturally written in terms of superspace couplings involving $X_{NL}$.  

The Super-Higgs mechanism is of course not new, however this approach emphasises how the Goldstino couples to general matter supercurrents, both through Goldberger-Treiman type derivative couplings and non-derivative couplings.  Whilst our effective superspace action does not have the full complexity of supergravity, including complicated expressions involving the K\"ahler metric, we gain a simpler more transparent construction that constrains the off-shell completion.  This may be useful for analysing Goldstino couplings to $U(1)_R$ currents and may be naturally extended to the use of $S^{\alpha \dot{\alpha}}$ and $R^{\alpha \dot{\alpha}}$ current multiplets \cite{Komargodski:2010rb}. It may also offer an in principle phenomenological way to determine the correct supercurrent multiplet that describes nature.  Furthermore, this setup allows for straightforward analysis of models with many Goldstini \cite{Cheung:2010mc} and also with Pseudo-Goldstini \cite{Argurio:2011hs,Argurio:2011gu}.

To find these \emph{new} superspace terms that we must add, we take suitable couplings of the Goldstino multiplet $X_{NL}$, the supercurrent multiplet $\mathcal{J}^{\alpha \dot{\alpha}}$,  the Graviton multiplet $H^{\alpha \dot{\alpha}}$,   $M_{Pl}$ and constant term $\braket{x}+C$, constrained by dimensional analysis. This is reasonable in a weak field expansion in $1/M_{Pl}$.  For a theory without matter the procedure is straightforward.  
For a theory with matter, to demonstrate the full shift of the super-Higgs mechanism, the Goldberger-Treiman relation(s) should appear explicitly.  This component is related to various couplings of superfields to the Goldstino multiplet and are then accounted for \cite{Bagger:2007gm}.   We then demonstrate the super-Higgs mechanism with matter supercurrents.

The outline of this paper is as follows:  In the next section we will review the constrained Goldstino multiplet $X_{NL}$ and its relation to the Ferrara-Zumino (F-Z) \cite{Ferrara:1974pz} supercurrent multiplet $\mathcal{J}^{\alpha \dot{\alpha}}$.  We then compute the effective action in components that reproduces the terms necessary for the super-Higgs mechanism.  Next we promote the component terms to a full superspace effective action, by coupling the Goldstino multiplet to the supercurrent multiplet and determine that the superspace formulation of these new terms correctly reproduces the components of the super-Higgs mechanism.  In an appendix we include a review of \cite{Rychkov:2007uq}, which is the component formulation of the super-Higgs mechanism. We adopt two-component spinor notation throughout.

\section{Non-Linear SUSY coupled to Supergravity}\label{nonlinear}
To describe a Goldstino supermultiplet we may start from a left handed chiral superfield $\bar{D}_{\dot{\alpha}}X=0$  and apply the constraint $X^2=0$.  The solution to this equation is given by the nonlinear Goldstino multiplet \cite{Komargodski:2009rz,Dumitrescu:2011zz}
\be
X_{NL}= \frac{G^2}{2F}+\sqrt{2}\theta G+ \theta^2 F.
\ee
The field $G^{\alpha}$ is the Goldstino and $F$ the Auxiliary field. $f$ is the scale of supersymmetry breaking. In general one must integrate out $F$, which may be complicated to do in practice, but will lead as $ \big<F\big> = f+... $, where the ellipses may often be ignored due to terms with higher derivatives. There may also be an  additive constant, $C$, in the definition of the scalar component $x$. The Goldstino multiplet also satisfies the conservation equation \cite{Ferrara:1974pz}
\be
\bar{D}^{\dot{\alpha}}\mathcal{J}_{\alpha \dot{\alpha}}=D_{\alpha} X.
\ee
where $\mathcal{J}_{\alpha \dot{\alpha}}=-2\sigma^{\mu}_{\alpha\dot{\alpha}}J_{\mu}$ is the Ferrara-Zumino supercurrent multiplet, which is a real linear multiplet: $D^2J= \bar{D}^2 J=0$.  In components it is given by 
\bea
\mathcal{J}_{\mu}&=&j_{\mu}+\theta^{\alpha}(S_{ \mu \alpha }+\frac{1}{3}\sigma_{\mu \alpha\dot{\alpha}} \bar{\sigma}^{\nu \dot{\alpha} \beta}S_{\nu \beta})+ \frac{i}{2}\theta^2 \partial_{\mu}(\bar{x}+\bar{C})+ \bar{\theta}_{\dot{\alpha}}(\bar{S}_{\mu}^{\dot{\alpha}}+\frac{1}{3}(\epsilon^{\dot{\alpha}\dot{\gamma}}\bar{S}_{\nu\dot{\beta}}\bar{\sigma}^{\nu \dot{\beta}\alpha} \sigma_{\mu \alpha \dot{\gamma}} )) \nonumber \\&- & \frac{i}{2}\bar{\theta}^2 \partial_{\mu}(x+C) +\theta^{\alpha} \sigma^{\nu}_{\alpha \dot{\alpha}}\bar{\theta}^{\dot{\alpha}}(2T_{\nu\mu}-\frac{2}{3}\eta_{\mu\nu} T-\frac{1}{4}\epsilon_{\nu\mu\rho \sigma}\partial^{[\rho}j^{\sigma]})+...\eea
$J_{\mu}$ is the R-current, $S^{\alpha}_{\mu}$ the supercurrent, $T_{\mu\nu}$ the stress energy tensor and we have made explicit the presence of a complex constant $C$ in the definition of the scalar component $x$, the term $\braket{x}+C$ will play the part of a cosmological constant as shown in the remainder of the paper. This $C$ may be equivalent to the auxiliary terms found in \cite{Weinberg:2000cr,Stelle:1978ye,Ferrara:1978em,Cremmer:1978hn} however from this perspective it appears in the scalar component of $X_{NL}$.  The term $\braket{x}=\frac{4}{3}\braket{G^2}$ is a Goldstino condensate and has been discussed before in \cite{Ellis:2011mz} (the coefficient appearing here was derived in \cite{Komargodski:2009rz}).  The ellipses may be determined from a shift from $ y^{\mu} =x^{\mu}-i\theta^{\alpha} \sigma^{\mu}_{\alpha\dot{\alpha}} \bar{\theta}^{\dot{\beta}}$. Our metric conventions are mostly minus, $\eta_{\mu\nu}=(1,-1,-1,-1)$. The supersymmetry current algebra is 
\be\label{algebra} 
\{\bar{Q}_{\dot{\alpha}},S_{\alpha \mu}\}= \sigma^{\nu}_{\alpha \dot{\alpha}}(2T_{\mu\nu}+i\partial_{\nu}j_{\mu}-i\eta_{\mu\nu}\partial^{\lambda}j_{\lambda}-\frac{1}{4}\epsilon_{\nu\mu\rho \lambda}\partial^{\rho}j^{\lambda})
\ee
\be
\{Q_\beta, S_{\mu\alpha}\}=2i\epsilon_{\lambda\beta}(\sigma_{\mu\rho})^\lambda_\alpha\partial^\rho(\bar{x}+\bar{C})
\ee
where the first term of \ref{algebra} is the conserved symmetric energy tensor and the remaining terms are Schwinger Terms that vanish in the vacuum. It is straightforward to use the definition of the Supercharge, $Q_{\alpha}=\int \! \! d^3x  S^{0}_{\alpha}$, to relate this expression to the super algebra 
\be
\{Q_{\alpha},\bar{Q}_{\dot{\alpha}}\}=2\sigma^{\mu}_{\alpha\dot{\alpha}}P_{\mu}.\textit{}
\ee
\newline
The simplest non linear Goldstino superfield $X_{NL}$ action that breaks supersymmetry spontaneously is given by 
\be
\mathcal{L}= \int d^4 \theta X^{\dagger}_{NL} X_{NL} + \int d^2\theta f^{\dagger} X_{NL}+\int d^2 \bar{\theta} fX^{\dagger}_{NL}. \label{Goldstinoaction}
\ee
As shown in \cite{Komargodski:2009rz} this action, in the absence of other couplings involving the Auxiliary field, is equivalent to the full A-V action in components
\be
\mathcal{L}_{AV}=-|f|^2+i\partial_\mu\bar{G}\bar{\sigma}^\mu G+\frac{1}{4|f|^2}\bar{G}^2\partial^2G^2-\frac{1}{16|f|^6}G^2\bar{G}^2\partial^2G^2\partial^2\bar{G}^2. \label{AVaction}
\ee
Let us for the moment focus on only the first two terms, which comprise the A-V action, ignoring terms with a higher number of derivatives.  The relevant terms in the A-V action are the supersymmetry breaking term and the Goldstino kinetic terms.   The supersymmetry breaking term $-|f|^2$ in the action is a cosmological constant.  The minimum of the scalar potential is $V_{\text{min}}= +|f|^2$ which is positive definite, and we see that global supersymmetry is broken. The supergravity action will also  generate a cosmological constant, but of opposite sign.  If these are made to be equal, the overall cosmological constant vanishes.  This will appear shortly. \newline

We introduce the linear supergravity action which provides kinetic terms for the Gravitino.  The supergravity fields are embedded in a real vector superfield. In Wess-Zumino gauge the components are given by 
\be 
H_{\mu}= \theta^{\alpha} \sigma^{\nu}_{\alpha \dot{\alpha}}\bar{\theta}^{\dot{\alpha}}(h_{\mu\nu}-\eta_{\mu\nu}h)+ \bar{\theta}^2 \theta^{\alpha}(\psi_{\mu \alpha}+ \sigma_{\mu \alpha \dot{\alpha}}\bar{\sigma}^{\rho\dot{\alpha} \beta}\psi_{\rho\beta})\nonumber \ee 
\be+\frac{i}{2}\theta^2M_\mu-\frac{i}{2}\bar{\theta}^2M^{\dagger}_\mu+ \theta^2 \bar{\theta}_{\dot{\alpha}}(\bar{\psi}_{\mu }^{\dot{\alpha}}+ \bar{\sigma}_{\mu}^{\dot{\alpha}\alpha}\sigma^{\rho}_{\alpha \dot{\beta}}\bar{\psi}_{\rho}^{\dot{\beta}}) - \frac{1}{2}\theta^2\bar{\theta}^2 A_{\mu}
\ee
where $H_{\mu}=\frac{1}{4}\bar{\sigma}_\mu^{\dot{\alpha}\alpha} H_{\alpha\dot{\alpha}}$, $h_{\mu\nu}$ is the linear Graviton, $\psi_\mu^\alpha$ is the Gravitino and $M_\mu, A_\mu$ are Auxiliary fields. The kinetic terms of the supergravity action are given by \cite{Dumitrescu:2011zz} 
\be
-\int d^4 \theta H^{\mu} E_{\mu}^{FZ} =  -\frac{1}{2}\epsilon^{\mu\nu \rho \sigma}(\psi_{\mu\alpha}\bar{\sigma}_{\nu}^{\dot{\alpha}\alpha}\partial_{\rho}\bar{\psi}_{\sigma \dot{\alpha}}-\bar{\psi}_{\mu\dot{\alpha}}\bar{\sigma}_{\nu}^{\dot{\alpha}\alpha}\partial_{\rho}\psi_{\sigma \alpha})
-\frac{1}{3}|\partial_\mu M^\mu|^2+\frac{1}{3}A_\mu A^\mu... \label{gravitykinetic}\ee
The ellipses include a linearised Graviton kinetic term.  $E_{\alpha \dot{\alpha}}^{FZ}$  is defined as
\be 
E_{\alpha \dot{\alpha}}^{FZ}=\bar{D}_{\dot{\tau}}D^2\bar{D}^{\dot{\tau}}H_{\alpha \dot{\alpha}} +\bar{D}_{\dot{\tau}}D^2\bar{D}_{\dot{\alpha}}H_{\alpha}^{ \dot{\tau}} +D^{\gamma}\bar{D}^2 D_{\alpha}H_{\gamma\dot{\alpha}} -2\partial_{\alpha \dot{\alpha}}\partial^{\gamma \dot{\tau}}H_{\gamma \dot{\tau}} .
\ee
This gives a kinetic term for the Graviton and Gravitino but the remaining fields are Auxilliary and therefore not dynamical. The Auxiliary field $A_\mu$ integrates out to give $A_\mu= \frac{j_\mu}{M_{Pl}}+...$, with the ellipses denoting higher order terms in $1/F$ and $1/\bar{M}_{Pl}$.  The complex field  $M_\mu$ plays a role in generating a cosmological constant once we weakly couple the supercurrent mutliplet to linear supergravity. \newline

Now that we have introduced the Goldstino and supergravity actions, we would like to couple the supercurrent multiplet to a linear supergravity muliplet \cite{Dumitrescu:2011zz}
\be\label{JH1} 
\frac{1}{8\bar{M}_{Pl}} \int d^4 \theta \mathcal{J}_{\alpha\alphad}H^{\alpha\alphad}=  \frac{1}{2\bar{M}_{Pl}}(h_{\mu \nu}T^{\mu \nu } + \psi_{\mu}S^{\mu}+\bar{\psi}_\mu\bar{S}^\mu- j^{\mu}A_{\mu}) \nonumber \ee 
\be -\frac{1}{4\bar{M}_{Pl}}\partial_\mu M^\mu (x+C) 
 -\frac{1}{4\bar{M}_{Pl}}\partial_\mu M^{\dagger\mu}(\bar{x}+\bar{C}) \label{currentcoupling}
\ee  
As we have integrated by parts on $\partial_{\mu} C$ there is also a total derivative term.  Also $\bar{M}_{pl}=M_{pl}/8\pi$ is the reduced Planck mass, $M_{pl}=(8\pi G)^{-\frac{1}{2}}$ where $G$ is the Gravitational constant. It is useful here to recall that the supersymmetry current multiplet $J^{\alpha\dot{\alpha}}$ contains at order $\theta^{\beta}$ a contribution from fermionic matter and a term proportional to the supersymmetry breaking term,
\be
S^\mu= S^\mu_{\text{matter}} +i\sqrt{2} f\sigma^{\mu}_{\alpha \dot{\alpha}} \bar{G}^{\dot{\alpha}} \ \ , \ \  \bar{S}^{\mu \dot{\alpha}}= \bar{S}^\mu_{\text{matter}} +i\sqrt{2}f^{\dagger} \bar{\sigma}^{\mu \dot{\alpha}\alpha} G_{\alpha}.
\ee
\section{Super-Higgs as an Effective Action}
Before we write down a superspace effective action, whose components reproduce the super-Higgs mechanism, it is worthwhile asking if one can indeed generate the necessary terms radiatively.  In this section we focus on the component formalism and explicitly compute the relevant terms in the effective action.  We shall find that the necessary terms all arise from couplings to the cosmological constant terms that we introduced in our definition of the supercurrent multiplet. 

\subsection{The cosmological constant}
First we focus on the cosmological constant.  Varying the combination of \ref{JH1} and \ref{gravitykinetic} with respect to $M^\mu$ (treating $M_\mu$ and $M_\mu^{\dagger}$ as independent fields) we find that
\be
\partial_\nu\left(\frac{1}{3}\partial_\mu M^{\dagger\mu}+\frac{1}{4\bar{M}_{Pl}}(x+C)\right)=0
\ee

\noindent which has as a solution

\be\label{solution}
M^{*}=\partial_\mu M^{\mu\dagger} = -\frac{3}{4\bar{M}_{Pl}}(x+C)+D
\ee

\noindent where $D$ is an arbitrary mass dimension 2 integration constant \cite{Ogievetsky:1980qp} which for the purposes of our discussion can be set to zero. Substituting this result back in the original action, and integrating out $A^{\mu}$, gives the terms
\be\label{exp3}
\mathcal{L}_{\text{aux}}=\frac{3(x+C)(\bar{x}+\bar{C})}{16\bar{M}_{Pl}^2}-\frac{3j_{\mu}j^{\mu}}{16\bar{M}^2_{Pl}}.
\ee

We can see that this mechanism provides a non-vanishing contribution to the vacuum energy density of the form \cite{Weinberg:2000cr,Ferrara:1978rk}

\be
\rho_{\text{Vac}} = -\frac{3|C|^2}{16\bar{M}_{Pl}^2}-\frac{3\big{<}x\bar{x}\big{>}}{16\bar{M}_{Pl}^2}-\frac{3C\big{<}\bar{x}\big{>}}{16\bar{M}_{Pl}^2}-\frac{3\bar{C}\big{<}x\big{>}}{16\bar{M}_{Pl}^2}
\ee 

\noindent which upon using $\big<x\big>=\big<\bar{x}\big>=\big<x\bar{x}\big>=0$ simplifies to,

\be
\rho_{\text{Vac}} = -\frac{3|C|^2}{16\bar{M}_{Pl}^2}.\ee


Taking the cosmological constant $|f|^2$ from the Goldstino action we may cancel the overall cosmological constant by setting 
\be\label{canc}
\frac{3|C|^2}{16\bar{M}_{Pl}^2}=|f|^2,
\ee 
and hence 
\be\label{cres} 
C = \frac{4}{\sqrt{3}}\bar{M}_{Pl}f.
\ee
In general $\braket{x}=\frac{4}{3}\braket{G^2}$ is not zero and therefore one would find
\be
 B=\braket{x}+|C|= \frac{4}{\sqrt{3}}\bar{M}_{Pl}|f|.
\ee
It has been shown in \cite{Deser:1977uq} that whether one generates a Gravitino mass, or not, depends on  whether these terms cancel, which is related to whether we are in Minkowski background or anti de Sitter space.  We take them to exactly cancel such that the massive Gravitino does appear.

At this order in $\bar{M}_{Pl}$ there is a contribution to the Goldstino self coupling coming from \ref{exp3}, this takes the form
\be
\mathcal{L}_{\text{aux}} \supset \frac{3(C\bar{x}+\bar{C}x)}{16\bar{M}^2_{Pl}} = m_{3/2}\bar{G}^2+m^{\dagger}_{3/2}G^2.
\ee


\subsection{The component terms}
Starting from \refe{currentcoupling} it is straightforward to see that the effective action generated from linear supergravity contains
\be
\mathcal{L}_{\text{eff}}\supset \frac{i}{8M^2_{pl}} (\tilde{G}_{3/2}(0) \psi^{\alpha}_{\mu} (\sigma^{\mu\nu})_{\alpha}^{\beta}\psi_{\beta\nu}+\tilde{\bar{G}}_{3/2}(0)  \bar{\psi}_{\dot{\alpha}\mu} (\bar{\sigma}^{\mu\nu})^{\dot{\alpha}}_{\dot{\beta}}\bar{\psi}^{\dot{\beta}}_{\nu}+S^{\alpha}_{\mu} \tilde{D}^{\mu\nu}_{\alpha\beta}(0)S^{\nu\beta}+ \bar{S}_{\dot{\alpha}}^{\mu} \tilde{\bar{D}}_{\mu\nu}^{\dot{\alpha}\dot{\beta}}(0)\bar{S}^{\nu}_{\dot{\beta}}). \label{effectiveactionterms}
\ee
We define the time ordered current correlators
\be
\braket{T[S^{\mu}_{\alpha}(p) S^{\nu}_{\beta}(-p) ] }=\tilde{\Pi}^{\mu\nu}_{\alpha \beta}\tilde{G}_{3/2}(p^2) 
\ee 
\be
\braket{T[S^{\mu}_{\alpha}(x) S^{\nu}_{\beta}(0)]  }=\Pi^{\mu\nu}_{\alpha \beta}G_{3/2}(x^2)
\ee
which are related by a Fourier Transform.  Additionally we define the two point function 
\be
\braket{\psi^{\mu}_{\alpha}(x)\psi^{\nu}_{\beta}(y)}= \int \frac{d^4 p}{(2\pi)^4}   \tilde{D}^{\mu\nu}_{\alpha\beta}(p^2) e^{-ip.(x-y)}
\ee
where we take that the two point function has the form 
\be
 \tilde{D}^{\mu\nu}_{\alpha\beta}(p^2)=-\frac{1}{3} \frac{im_{3/2}}{p^2-m^{2}_{3/2}}\left(\eta^{\mu\nu}\epsilon_{\alpha\beta}+\frac{2i}{3}(\epsilon\sigma^{\mu\nu})_{\alpha\beta}\right).
\ee
The mass pole
\be 
m_{3/2}=\frac{ i\tilde{G}_{3/2}(0)}{8M^2_{pl}}
\ee
is a consequence of a standard geometric sum of mass insertions. For the model outlined in this paper, as will be demonstrated below,
\be\label{g3}
 \tilde{G}_{3/2}(0)= 2B \ \ \ \ \text{and} \ \ \ \  \tilde{D}^{\mu\nu}_{\alpha\beta}(0)=-\frac{i}{3m_{3/2}}\left(\eta^{\mu\nu}\epsilon_{\alpha\beta}+\frac{2i}{3}(\epsilon\sigma^{\mu\nu})_{\alpha\beta}\right),
\ee
such that one recovers the effective contributions 
\be \label{g4}
\mathcal{L}_{\text{eff}}=-i\left(m_{3/2}\psi_\mu^\alpha(\sigma^{\mu\nu})_\alpha^\beta\psi_{\nu\beta}+ m^{\dagger}_{3/2} \bar{\psi}_{\mu\dot{\alpha}}(\bar{\sigma}^{\mu\nu})^{\dot{\alpha}}_{\dot{\beta}}\bar{\psi}_{\nu}^{\dot{\beta}}\right)\nonumber \ee 
\be -\frac{2f^2}{3M^2_{Pl}m_{3/2}}G^\alpha G_\alpha-\frac{2(f^{\dagger})^2}{3M^2_{Pl}m^{\dagger}_{3/2}}\bar{G}_{\alphad}\bar{G}^{\alphad}\nonumber
\ee 
\be   
+\frac{if^{\dagger}\sqrt{2}}{12\bar{M}^2_{Pl}m^{\dagger}_{3/2}} G^{\alpha} \sigma^{\mu}_{\alpha \dot{\alpha}}\bar{S}^{\dot{\alpha}}_{\mu\text{matter}}
+\frac{if\sqrt{2}}{12\bar{M}^2_{Pl}m_{3/2}}\bar{G}_{\dot{\alpha}} \bar{\sigma}^{\mu \dot{\alpha}\alpha}S_{\alpha \mu\text{matter}}.  
\ee
From these we see that the cosmological constant factor $\frac{B}{4\bar{M}_{Pl}^2}$ determines the gravitino mass $m_{3/2}$.  It is only after we require the vanishing cosmological constant, in the form of \ref{canc}, that we find the relation $m_{3/2}  =\frac{f}{\sqrt{3}\bar{M}_{Pl}}$.


Finally, we note that the  local supersymmetry transformations are also \emph{modified} by the presence of a non vanishing $B$, these become
\be
\delta \psi_{\mu\alpha} =-\left(2\bar{M}_{Pl}\partial_\mu\epsilon_\alpha+i\frac{B}{4\bar{M}_{Pl}}\sigma_{\mu\alpha\dot{\alpha}}\epsilon^{\dot{\alpha}}\right).
\ee
Taking \refe{g4} and including also the Goldberger Trieman relations one then has all the necessary terms to apply the shifts \refe{shifts1} and obtain the massive Gravitino action coupled to matter supercurrents \refe{massiveWeylcoupled}.  However we will defer this until we have \emph{rederived} these very same terms from effective superspace terms, in the next section.  There are also corrections of order $p^2/m^2_{3/2}$   derived from the effective action by integrating out the Gravitino propagator \refe{g3}. 
\subsubsection{Computing the current correlator}
Here we demonstrate the evaluation of the time ordered current correlator appearing in \ref{effectiveactionterms}
\be 
\braket{T [S^{\mu}_{\alpha}(x) S^{\nu}_{\beta}(0) ] }.
\ee
We wish to write this correlator in terms of the super algebra.  To do this we take first the definition 
\be  
\braket{T [S^{\mu}_{\alpha}(x) S^{\nu}_{\beta}(y) ] }=\theta(x^{0}-y^{0})\braket{S^{\mu}_{\alpha}(x) S^{\nu}_{\beta}(y)  }-\theta(y^{0}-x^{0})\braket{ S^{\nu}_{\beta}(y) S^{\mu}_{\alpha}(x) },
\ee
where $\theta(x^0-y^0)$ is the Heaviside function and the minus sign between the two terms on the right hand side is standard for fermionic operators. 
Using the standard manipulation
\be  
\partial_{\mu}\braket{T [S^{\mu}_{\alpha}(x) S^{\nu}_{\beta}(y) ] }=\delta(x^{0}-y^{0})\braket{ \{S^{0}_{\alpha}(x) ,S^{\nu}_{\beta}(y)\}  }+\braket{ T[\partial_{\mu}S^{\mu}_{\beta}(x) S^{\nu}_{\alpha}(y)] }
\ee
and then integrating by parts
\be 
-y^{\rho}\partial_{\mu}\braket{T [S^{\mu}_{\alpha}(x) S^{\nu}_{\beta}(y) ] }=(\partial_{\mu}y^{\rho}) \braket{T [S^{\mu}_{\alpha}(x) S^{\nu}_{\beta}(y) ] }
\ee
with $y^{\rho}$ as a four-vector, one obtains
\be
\braket{T [S^{\rho}_{\alpha}(x) S^{\nu}_{\beta}(y) ] }=- y^{\rho}\left(\delta(x^{0}-y^{0})\braket{ \{S^{0}_{\alpha}(x) ,S^{\nu}_{\beta}(y)\}  }+\braket{ T[\partial_{\mu}S^{\mu}_{\beta}(x) S^{\nu}_{\alpha}(y)] } \right).
\ee 
Using  $\partial_{\mu}S^{\mu}=0$ to remove the second term then $\int d^3x S^0_{\alpha}(x)=Q_{\alpha}$ gives 
\be 
\int d^4 x\braket{T [S^{\rho}_{\alpha}(x) S^{\nu}_{\beta}(y) ] }=-y^{\rho}  \braket{ \{ Q_{\alpha} ,S^{\nu}_{\beta}(y)\}  }.
\ee
Inserting
\be
\{Q_\beta, S_{\mu\alpha}\}=2i\epsilon_{\lambda\beta}(\sigma_{\mu\rho})^\lambda_\alpha\partial^\rho(\bar{x}+\bar{C})
\ee
integrating by parts, we find
\be 
\int d^4 x\braket{T [S^{\rho \alpha}(x) S^{\nu \beta}(y) ] }= (\partial^{\sigma}  y_{\rho}  )2i\epsilon_{\lambda\alpha}(\sigma_{\nu\sigma})^\lambda_\beta (\bar{x}+\bar{C}).
\ee
Thus we see that the constant $\braket{x}+C$ appears in the current correlator.  We mention that if one instead evaluates the sum rule for $\{\bar{Q}_{\dot{\alpha}},S^{\mu}_{\alpha}(x)\}  $  then one may show the massless Goldstino pole \cite{Salam:1974zb,Witten:1981nf}.

  Digressing slightly, we make the comment that both $f^2$ and $\braket{x}+C$ are contained within the current algebra and at this point are both free parameters.  However the requirement of vanishing overall cosmological constant suggests that there is really only one free parameter. One might wish for some sort of Ward identity which tunes the value of $C$ to preserve local SUSY. It is also tempting to suggest that whatever mechanism gives rise to supersymmetry breaking should generate both terms.  There are at least two independent contributions to   $\braket{x}+C$ (perhaps more?).  It may also be worthwhile to speculate on what types of symmetries $X_{NL}$ may be forced to obey such that the vev of the scalar component and the vev of the $F$ term are related to cancel the overall cosmological constant.

 \section{The Superspace Terms}
In the previous section we demonstrated that certain terms needed for the super-Higgs mechanism are generated in an effective action that combines linear supergravity and the Goldstino multiplet.  In this section we will promote the effective terms in components to full superspace terms in the action.  We will demonstrate that the components combined do indeed satisfy the super-Higgs mechanism.   Our starting point of this superspace effective action is to assume all the relevant pieces of section \ref{nonlinear}. In particular we also assume the vanishing overall cosmological constant \refe{canc}. Next we introduce a \emph{new} term, we add to the supergravity action 
\be 
-\frac{1}{64}\int d^4 \theta \left(\frac{m_{3/2}X_{NL}}{f}+\frac{m^{\dagger}_{3/2}X^\dagger_{NL}}{f^{\dagger}}\right)\left[\bar{D}_{\dot{\alpha}} D_{\alpha} H^{\beta \dot{\beta}}\bar{D}^{\dot{\alpha}} D^{\alpha} H_{\beta \dot{\beta}}-\frac{4}{3}(\bar{D}_{\dot{\alpha}} D_{\alpha} H^{\alpha \dot{\alpha}})^2\right] \label{mass12}  
\ee
which in components gives a Gravitino self coupling 
\be
-i\left(m_{3/2}\psi_{\mu}^{ \alpha} ( \sigma^{\mu\nu})_{\alpha}^{\beta}  \psi_{\nu  \beta} +m_{3/2}^{\dagger} \bar{\psi}_{\mu  \dot{\alpha}} ( \bar{\sigma}^{\mu\nu})^{\dot{\alpha}}_{\dot{\beta}} \bar{\psi}_{\nu }^{ \dot{\beta}}\right)+... 
\ee
The second superspace term was introduced to remove terms of the form $\psi^2$. The ellipses denote some higher order terms or derivative couplings, which we will ignore.  Different off-shell completions will generate different couplings at this order, including couplings between the Goldstino and the off-shell supergravity fields, which may be interesting to catalogue but deviate too far from the present discussion.  \newline

Next we introduce the \emph{new} superspace term 
\be 
\frac{9}{784}\int d^4 \theta \left(\frac{m_{3/2}X_{NL}}{f}+\frac{m^{\dagger}_{3/2}X^\dagger_{NL}}{f^{\dagger}}\right)  \frac{\mathcal{J}_{\alpha \dot{\alpha}} \mathcal{J}^{\alpha \dot{\alpha}}}{|f|^2}, \label{JJ2} 
\ee
which introduces the Goldstino couplings 
\be
\frac{i}{2\sqrt{6}\bar{M}_{Pl}} (G^{\alpha} \sigma^{\mu}_{\alpha \dot{\alpha}}\bar{S}^{\dot{\alpha}}_{\mu\text{matter}}
+\bar{G}_{\dot{\alpha}} \bar{\sigma}^{\mu \dot{\alpha}\alpha}S_{\alpha \mu\text{matter}}) -2m_{3/2}\bar{G}_{\dot{\alpha}}\bar{G}^{\dot{\alpha}} -2m^{\dagger}_{3/2}G^{\alpha}G_{\alpha}. \label{funny}
\ee
\noindent There are also some higher order terms, mostly suppressed by higher orders in $1/F$.  
\newline
\noindent Collecting the relevant terms one obtains
\be
\mathcal{L}=-\frac{1}{2}\epsilon^{\mu\nu \rho \sigma}(\psi_{\mu\alpha}\bar{\sigma}_{\nu}^{\dot{\alpha}\alpha}\partial_{\rho}\bar{\psi}_{\sigma \dot{\alpha}}-\bar{\psi}_{\mu\dot{\alpha}}\bar{\sigma}_{\nu}^{\dot{\alpha}\alpha}\partial_{\rho}\psi_{\sigma \alpha})-i\left( m_{3/2}\psi_{\mu}^{ \alpha} ( \sigma^{\mu\nu})_{\alpha}^{\beta}  \psi_{\nu  \beta} +m_{3/2}^{\dagger} \bar{\psi}_{\mu  \dot{\alpha}} ( \bar{\sigma}^{\mu\nu})^{\dot{\alpha}}_{\dot{\beta}}  \bar{\psi}_{\nu }^{ \dot{\beta}}\right)\nonumber
\ee
\be
+\frac{i}{2\sqrt{6}\bar{M}_{Pl}} (G^{\alpha} \sigma^{\mu}_{\alpha \dot{\alpha}}\bar{S}^{\dot{\alpha}}_{\mu\text{matter}}
+\bar{G}_{\dot{\alpha}} \bar{\sigma}^{\mu \dot{\alpha}\alpha}S_{\alpha \mu\text{matter}}) -m_{3/2}\bar{G}_{\dot{\alpha}}\bar{G}^{\dot{\alpha}}-m^{\dagger}_{3/2}G^{\alpha}G_{\alpha}
\nonumber
\ee
\be
+ \frac{1}{2\bar{M}_{Pl}}(\psi_{\mu}S^{\mu}+\bar{\psi}_\mu\bar{S}^\mu)+ \frac{1}{2}iG^{\alpha}\sigma^{\mu}_{\alpha \dot{\alpha}}\partial_{\mu}\bar{G}^{\dot{\alpha}}+\frac{1}{2}i \bar{G}_{\dot{\alpha}}\bar{\sigma}^{\mu\dot{\alpha}\alpha}\partial_{\mu}G_{\alpha}+...\label{mainlagrangian}
\ee

\subsection{The case of no matter}
We consider first the case of vanishing matter contributions, $S^\mu_{\alpha\text{matter}}=0$. The component Lagrangian \refe{mainlagrangian} reduces to
\be
\mathcal{L}=-\frac{1}{2}\epsilon^{\mu\nu \rho \sigma}(\psi_{\mu\alpha}\bar{\sigma}_{\nu}^{\dot{\alpha}\alpha}\partial_{\rho}\bar{\psi}_{\sigma \dot{\alpha}}-\bar{\psi}_{\mu\dot{\alpha}}\bar{\sigma}_{\nu}^{\dot{\alpha}\alpha}\partial_{\rho}\psi_{\sigma \alpha})-i\left( m_{3/2}\psi_{\mu}^{ \alpha} ( \sigma^{\mu\nu})_{\alpha}^{\beta}  \psi_{\nu  \beta} +m_{3/2}^{\dagger}\bar{\psi}_{\mu  \dot{\alpha}} ( \bar{\sigma}^{\mu\nu})^{\dot{\alpha}}_{\dot{\beta}}  \bar{\psi}_{\nu }^{ \dot{\beta}}\right)\nonumber
\ee
\be
-m_{3/2}\bar{G}_{\dot{\alpha}}\bar{G}^{\dot{\alpha}} -m^{\dagger}_{3/2}G^{\alpha}G_{\alpha}
+\frac{i}{\sqrt{2}\bar{M}_{Pl}}(F \psi^{\alpha}_{\mu} \sigma^{\mu}_{\alpha \dot{\alpha}} \bar{G}^{\dot{\alpha}}+F^{\dagger}\bar{\psi}_{\mu\dot{\alpha}} \bar{\sigma}^{\dot{\alpha}\alpha}G_{\alpha}) \nonumber
\ee
\be
+ \frac{1}{2}iG^{\alpha}\sigma^{\mu}_{\alpha \dot{\alpha}}\partial_{\mu}\bar{G}^{\dot{\alpha}}+\frac{1}{2}i \bar{G}_{\dot{\alpha}}\bar{\sigma}^{\dot{\alpha}\alpha}\partial_{\mu}G_{\alpha}+...
\ee
In this case, we can realise the super-Higgs mechanism by applying the shifts \cite{Cremmer:1978iv}
\be
\Psi_{\mu \alpha}= \psi_{\mu \alpha}- \frac{i}{\sqrt{6}}\sigma_{\mu \alpha \dot{\alpha}}\bar{G}^{\dot{\alpha}}-\sqrt{\frac{2}{3}}\frac{\partial_\mu G_\alpha}{m_{\frac{3}{2}}} \ \ , \ \ \bar{\Psi}_{\mu}^{ \dot{\alpha}}= \bar{\psi}_{\mu}^{\dot{ \alpha}}-\frac{i}{\sqrt{6}}\bar{\sigma}_{\mu}^{\dot{\alpha}\alpha} G_{\alpha}-\sqrt{\frac{2}{3}}\frac{\partial_\mu \bar{G}^{\dot{\alpha}}}{m^{\dagger}_{\frac{3}{2}}} \label{shifts1},
\ee
then one reproduces 
\be
\mathcal{L}=
-\frac{1}{2}\epsilon^{\mu\nu \rho \sigma}(\Psi_{\mu}^{\alpha}\sigma_{\nu\alpha\dot{\alpha}}\partial_{\rho}\bar{\Psi}_{\sigma}^{\dot{\alpha}}-\bar{\Psi}_{\mu\dot{\alpha}}\bar{\sigma}_{\nu}^{\dot{\alpha}\alpha}\partial_{\rho}\Psi_{\sigma \alpha})-i\left(m_{3/2}\Psi_{\mu}^{ \alpha} ( \sigma^{\mu\nu})_{\alpha}^{\beta}  \Psi_{\nu  \beta} +m_{3/2}^{\dagger}\bar{\Psi}_{\mu  \dot{\alpha}} ( \bar{\sigma}^{\mu\nu})^{\dot{\alpha}}_{\dot{\beta}}\bar{\Psi}_{\nu}^{\dot{\beta}}\right)\nonumber
\ee
the Lagrangian of a massive Gravitino Weyl spinor $\Psi_{\mu}^{\alpha}$.

\subsection{Coupling to matter}
To couple the massive Gravitino to the matter supercurrent requires that all components of the shift of \refe{shifts1} couple to the matter supercurrent. After integration by parts, one of these shifted terms is the Goldberger Treiman relation. These terms are the most useful for phenomenology \cite{Bolz:2000fu}.  In the superspace formalism, the Goldberger Treiman terms appear in components in their non derivative form \cite{Bagger:2007gm,Argurio:2011hs} from superspace terms such as 
\be 
\int d^2 \theta \frac{ m_{\lambda}}{2f}X_{NL} W^{\alpha}W_{\alpha} +\int d^2 \bar{\theta} \frac{ m^{\dagger}_{\lambda}}{2f^{\dagger}}X^{\dagger}_{NL} \bar{W}_{\dot{\alpha}}\bar{W}^{\dot{\alpha}} \label{coup1}
\ee
and
\be 
\int d^4 \theta \frac{m^2_{0}}{|f|^2} X^{\dagger}_{NL}X_{NL} \Phi^{\dagger}\Phi   \label{coup2}
\ee 
where $\Phi$ represents some generic matter chiral superfield and $W^{\alpha}$ is the superfield strength tensor. $m_{\lambda}$ and $m_{0}$ are the soft supersymmetry breaking masses of Gauginos and Scalars. Expanding out these terms in superspace, first give the soft breaking mass terms and at linear order in the Goldstino, give the Golberger Treiman relations.  After collecting the components and use of the equation of motion, this will supply 
\be
\frac{G^{\alpha}}{\sqrt{2}f}\partial_{\mu}S^{\mu}_{\alpha \text{matter}}+\frac{\bar{G}_{\dot{\alpha}}}{\sqrt{2}f^{\dagger}}\partial_{\mu}\bar{S}^{\mu \dot{\alpha}}_{\text{matter}}.
\ee
Including these terms with the action given by \refe{mainlagrangian} and applying the shifts of \refe{shifts1} we find 
\be
\mathcal{L}= \frac{1}{2}\epsilon^{\mu\nu \rho \sigma}(\Psi_{\mu}^{\alpha}\sigma_{\nu\alpha\dot{\alpha}}\partial_{\rho}\bar{\Psi}_{\sigma}^{\dot{\alpha}}  -\bar{\Psi}_{\mu\dot{\alpha}}\bar{\sigma}_{\nu}^{\dot{\alpha}\alpha}\partial_{\rho}\Psi_{\sigma \alpha})
-i\left(m_{3/2}\Psi_\mu^\alpha(\sigma^{\mu\nu})_\alpha^\beta\Psi_{\nu\beta}+m_{3/2}^{\dagger}\bar{\Psi}_{\mu\dot{\alpha}}(\bar{\sigma}^{\mu\nu})^{\dot{\alpha}}_{\dot{\beta}}\bar{\Psi}_{\nu}^{\dot{\beta}}\right)\nonumber
\ee
\be+\frac{1}{2\bar{M}_{Pl}}(\Psi_{\mu}^\alpha (S^{\mu}_{\alpha})_\text{matter}+\bar{\Psi}_{\mu\alphad}(S^{\mu\alphad})_\text{matter})+...\label{massiveWeylcoupled}
\ee
The Lagrangian of a massive Weyl Gravitino coupled to matter.


\section{Applications}
So far what we have written is rather abstract.  In this section we wish to demonstrate some of the uses that an effective action of the super-Higgs mechanism in superspace may have.  To demonstrate this we first write down the effective action which, in components, will reproduce the linearised action that correctly describes the theory of multiple Goldstini \cite{Cheung:2010mc}. 

\subsection{Goldstini}
As a simple application, we would like to write an action whose components naturally reproduce the effect of multiple supersymmetry breaking sectors, and hence, multiple Goldstini. We use an index $i$ running from $1$ to $N$, to label the supersymmetry breaking sectors, with $F_{i}$ F-terms and $X_{i}$ Chiral superfields.  This is the interaction basis $\eta_{i}$.  The mass basis we will label $\{G,\zeta_{a} \}$. The notation is therefore no sum over the index $i$ appearing in $S_{i\mu}^\alpha$.  We define the Chiral superfields
\be
X_{iNL} = \frac{\eta_i^2}{2F_i}+\sqrt{2}\theta\eta_i+\theta^2 F_i
\ee
from which the components of the supercurrent may be found
\be
S_{\mu\text{total} }^\alpha = S_{\mu\text{matter}}^{\alpha}+\sum_{i}i\sqrt{2}F_i\sigma_{\mu\alpha\betad}\bar{\eta}^{\betad}_i=S_{\mu\text{matter}}^{\alpha}+i\sqrt{2}f_{eff}\sigma_{\mu\alpha\betad}\bar{G}^{\betad}.\label{newcurrent}
\ee
It is straightforward to see that the uneaten Goldstino do not appear in the supercurrent, which is a simple application of 
\be 
G^{\alpha}=\frac{1}{f_{eff}}\sum_{i}F_{i}\eta^{\alpha}_i
\ee
Where $f_{eff}^2$ is the sum of the $f^2_i$.   We now introduce the  $N$ Goldstino multiplet Lagrangian 
\be
\sum_{i} \int d^4\theta X_{i NL}X^{\dagger}_{i NL}+\left(\int d^2\theta f^{\dagger}_i X_{ i NL}+\int d^2\bar{\theta} f_i X^{\dagger}_{ i NL}\right).\ee
 The current supermultiplet to supergravity coupling is modified 
\be 
\frac{1}{M_{Pl}}\int d^4 \theta J_{\alpha\dot{\alpha}}H^{\alpha\dot{\alpha}} \supset   -\frac{1}{4\bar{M}_{Pl}} \partial_{\mu}M^{\mu}(x_i + C)
\ee
which after integrating out $\partial_{\mu}M^{\mu}$ and using $x_i = \frac{4}{3}\eta_i^2$ therefore leads to 
\be
\sum_i \frac{3(\frac{4}{3}\eta_{i}^2+C)(\frac{4}{3}\bar{\eta}^2_{i}+\bar{C})}{16M_{Pl}}\rightarrow \sum_i m^{\dagger}_{3/2}\eta^2_i+m_{3/2}\bar{\eta}^2_{i}
\ee
which are the uneaten Goldstini masses and a contribution to the Gravitino mass.  Once the overall cosmological constant is assumed  to vanish  $C= \frac{4}{\sqrt{3}}f_{eff}\bar{M}_{Pl}$,   this sets $m_{3/2}=f_{eff}/\sqrt{3}M_{pl}$.  this also fixes the masses of the other $\zeta_i$ Goldstini which are proportional to C   (and not on their respective $F_i$ as one might naively think from an effective theory approach).  It is the vanishing of the overall cosmological constant that sets $m_a=2m_{3/2}$: this seems to be the most intuitive argument for the Goldstini mass formula.  

Let us now compute the effective action.  In fact no new computation is necessary: After applying \refe{newcurrent} will reproduce exactly the same terms as found in \refe{g4}. Importantly, the uneaten Goldstini do not appear.

It is interesting to ask what may be learned about multiple Goldstini coupling to supersymmetric standard model matter.  The Goldstini obey analogue Goldberger Treiman type couplings from a natural extension of  \refe{coup1} and \refe{coup2}  and these have already been explored \cite{Cheung:2010mc}.   However, interestingly the goldstino to matter coupling of the form $\eta^\alpha \sigma^\mu_{\alpha\betad}\bar{S}^{\betad}_\mu/M_{Pl}$ only appears for the true Goldstino $G_\alpha$ and not for the Goldstini.

\section{Discussion}
In this paper we have written an effective action, in superspace, that manifestly respects global supersymmetry and whose components reproduce the super-Higgs mechanism.  The coefficients of these superspace terms appear to be chosen by hand to reproduce the necessary components.  We  demonstrate that these coefficients arise from an effective action.  After using this choice of coefficients, the components respect the necessary modified local supersymmetry transformations.  It is perhaps unfortunate that local supersymmetry does not seem to fix the coefficients at the level of superfields but suggest that we should interpret our action as an effective one, in any case. 

Still we think this setup is useful as it achieves the super-Higgs mechanism of the Goldstino multiplet and more interestingly, through the use of the supercurrent multiplet.  Additionally we have outlined how an effective action may be written that reproduces the results of multiple Goldstini. \cite{Cheung:2010mc}.

\paragraph{Acknowledgements} 
MM is funded by the Alexander Von Humboldt Foundation. We would like to thank Andreas Weiler, Daniel C.Thompson, Alberto Mariotti, Omer Gurdogan and Robert Mooney for interesting discussions. During part of this work G.T. was funded by an EPSRC studentship.

\appendix

\section{The Super-Higgs Mechanism: A Review}\label{review}
In this section we review the super-Higgs mechanism following closely the appendix of \cite{Rychkov:2007uq}.  We choose to review this appendix for two principle reasons: first it makes apparent the importance of couplings to the supercurrent $S^{\mu}_{\alpha}$, for example the term \refe{surprising},  and secondly because we wish to connect the above work directly with phenomenology and observation.

In two component spinor notation, if supersymmetry is broken by an F term vacuum expectation value, then the Goldstino must transform as $\delta_\epsilon \chi = \sqrt{2}F \epsilon$  and $\delta_\epsilon \bar{\chi} = \sqrt{2}F^{\dagger} \bar{\epsilon}$ where ($\epsilon,\bar{\epsilon}$) are  supersymmetry transformation parameters and $\sqrt{2}$ is a convention.  Treating supersymmetry as a global symmetry, Noether's theorem leads to a conserved supercurrent
\be 
\delta \mathcal{L}=(\partial_\mu \epsilon^{\alpha})S^\mu_{\alpha}+(\partial_\mu \bar{\epsilon}_{\dot{\alpha}})\bar{S}^{\mu \dot{\alpha}}= 0.
\ee
Integrating by parts one finds the variation of the action 
\be \label{variation}
\delta \mathcal{S}=-\int d^4x\left[  \epsilon^{\alpha}(\partial_{\mu} S^\mu_{\alpha})+ \bar{\epsilon}_{\dot{\alpha}}(\partial_{\mu} \bar{S}^{\mu \dot{\alpha}})\right] = 0.
\ee
The action may be determined
\be\label{action22}
S_{GT}= \int d^4 x  \left[\frac{\chi^{\alpha}}{\sqrt{2}F}\partial_{\mu}S^\mu_{\alpha}+\frac{\bar{\chi}_{\dot{\alpha}}}{\sqrt{2}F^{\dagger}}\partial_{\mu} \bar{S}^{\mu \dot{\alpha}}\right].
\ee
This gives the familiar Goldberger Treiman relation and a kinetic term for the Goldstino. The supercurrent should contain general matter contributions and a term proportional to the vev:
\be\label{ess}
S^{\mu}_{\alpha}= S^{\mu}_{\alpha \text{matter}} +i\sqrt{2} F\sigma^{\mu}_{\alpha \dot{\alpha}} \bar{\chi}^{\dot{\alpha}} \ \ , \ \  \bar{S}^{\mu \dot{\alpha}}= \bar{S}^{\mu \dot{\alpha}}_{\text{matter}} +i\sqrt{2}F^{\dagger} \bar{\sigma}^{\mu \dot{\alpha}\alpha} \chi_{\alpha}.
\ee
Invariance of the action under supersymmetry transformations implies the canonically normalised Goldstino kinetic terms
\be
\frac{1}{2}i \chi^{\alpha}\sigma^{\mu}_{\alpha \dot{\alpha}}\partial_{\mu}\bar{\chi}^{\dot{\alpha}}+\frac{1}{2}i \bar{\chi}_{\dot{\alpha}}\bar{\sigma}^{\mu\dot{\alpha}\alpha}\partial_{\mu}\chi_{\alpha},
\ee
these being related to each other by an integration by parts. The reader can verify that varying the above kinetic contribution with respect to $\chi^\alpha$ and $\bar{\chi}^{\alphad}$ independently will lead to the constraints $\partial_\mu S^{\mu\alpha}=0$ and $\partial_\mu \bar{S}^{\mu\alphad}=0$, as required in \ref{variation}, thus avoiding double counting the kinetic terms as would result by substituting \ref{ess} in \ref{action22} directly. \newline

We now introduce the Gravitino 
\be
S_{kin}= -\frac{1}{2}\int d^4x\epsilon^{\mu\nu \rho \sigma}(\psi_{\mu}^{\alpha}\sigma_{\nu\alpha\dot{\alpha}}\partial_{\rho}\bar{\psi}_{\sigma}^{\dot{\alpha}}-\bar{\psi}_{\mu\dot{\alpha}}\bar{\sigma}_{\nu}^{\dot{\alpha}\alpha}\partial_{\rho}\psi_{\sigma \alpha})
\ee
and consider weakly gauging gravity by the introduction of a Gravitino that couples to the supercurrent
\be
S_{\text{int} 1}= \int d^4 x \frac{1}{2\bar{M}_{Pl}} [\psi^{\alpha}_{\mu}S_{\alpha}^{\mu}+\bar{\psi}_{\dot{\alpha}\mu}\bar{S}^{\dot{\alpha}\mu}].
\ee
This term naturally leads to 
\be
\frac{iF}{\sqrt{2}\bar{M}_{Pl}}\psi^{\alpha}_{\mu}\sigma^{\mu}_{\alpha \dot{\alpha}}\bar{\chi}^{\dot{\alpha}} +\frac{iF^{\dagger}}{\sqrt{2}\bar{M}_{Pl}} \bar{\psi}_{\mu \dot{\alpha}}\bar{\sigma}^{\mu \dot{\alpha}\alpha}\chi_{\alpha},
\ee
Additionally one must introduce the term
\be\label{a0}
\mathcal{S}_{\text{int} 2}=\int d^4 x \frac{i}{2\sqrt{6}\bar{M}_{Pl}} [\chi^{\alpha} \sigma^{\mu}_{\alpha \dot{\alpha}}\bar{S}^{\dot{\alpha}}_{\mu}
+\bar{\chi}_{\dot{\alpha}} \bar{\sigma}^{\mu \dot{\alpha}\alpha}S_{\alpha \mu}],
\ee
to obtain 
\be\label{a54}
-\frac{2F^{\dagger}}{\sqrt{3}\bar{M}_{Pl}}\chi^{\alpha}\chi_{\alpha} -\frac{2F}{\sqrt{3}\bar{M}_{Pl}}\bar{\chi}_{\dot{\alpha}}  \bar{\chi}^{\dot{\alpha}} .
\ee
which is the first source of what will later become the Gravitino mass of the super-Higgs mechanism.  There is also a contribution from coupling directly to the cosmological constant to give the overall mass 
\be\label{a1}
-\frac{F^{\dagger}}{\sqrt{3}\bar{M}_{Pl}}\chi^{\alpha}\chi_{\alpha} -\frac{F}{\sqrt{3}\bar{M}_{Pl}}\bar{\chi}_{\dot{\alpha}}  \bar{\chi}^{\dot{\alpha}} .
\ee 
In addition, it generates a non derivative coupling between the Goldstino and the supercurrent
\be
\frac{i}{2\sqrt{6}\bar{M}_{Pl}}\bar{\chi}^{\dot{\alpha}}\sigma^\mu_{\alpha\dot{\alpha}}(S_\mu^{\alpha})_{\text{matter}}+\frac{i}{2\sqrt{6}\bar{M}_{Pl}}\chi^{\alpha} \sigma^{\mu}_{\alpha \dot{\alpha}}(\bar{S}^{\dot{\alpha}}_{\mu})_{\text{matter}}. \label{surprising}
\ee
In \cite{Rychkov:2007uq}, it was commented that it may seem surprising to add this new term \refe{surprising}, but that there is no contradiction as this new term vanishes when  $M_{Pl}\rightarrow \infty$.  It is therefore interesting to see that it arises \emph{quite} naturally after one computes the effective action \refe{effectiveactionterms} or instead from \refe{JJ2}.

To preserve local supersymmetry invariance under the \emph{modified} \cite{Deser:1977uq} transformations 
\bea
\delta\psi_{\mu\alpha}&=&-\bar{M}_{Pl}\left(2\partial_\mu\epsilon_\alpha+im_{\frac{3}{2}}\sigma_{\mu\alpha\dot{\alpha}}\epsilon^{\dot{\alpha}}\right)\\
\delta\chi_\alpha&=&\sqrt{2}F\epsilon_\alpha
\eea
one must add a Gravitino self-coupling term
\be
S_{m_{3/2}}=-i\int d^4 x\left(m_{3/2}\psi_\mu^\alpha(\sigma^{\mu\nu})_\alpha^\beta\psi_{\nu\beta}+m_{3/2}^{\dagger}\bar{\psi}_{\mu \dot{\alpha}}(\bar{\sigma}^{\mu\nu})^{\dot{\alpha}}_{\dot{\beta}}\bar{\psi}_{\nu}^{\dot{\beta}}\right),
\ee
\noindent provided that 
\be
m_{3/2}=\frac{F}{\sqrt{3}\bar{M}_{Pl}}.
\ee
\newline
\noindent Gathering all the terms together the overall Lagrangian is therefore
\be
\mathcal{L}= -\frac{1}{2}\epsilon^{\mu\nu \rho \sigma}
(\psi_{\mu}^{\alpha}\sigma_{\nu\alpha\dot{\alpha}}\partial_{\rho}\bar{\psi}_{\sigma}^{\dot{\alpha}}-\bar{\psi}_{\mu\dot{\alpha}}\bar{\sigma}_{\nu}^{\dot{\alpha}\alpha}\partial_{\rho}\psi_{\sigma \alpha})-i\left(m_{3/2}\psi_\mu^\alpha (\sigma^{\mu\nu})_\alpha^\beta \psi_{\nu\beta}+m_{3/2}^{\dagger}\bar{\psi}_{\mu \dot{\alpha}}(\bar{\sigma}^{\mu\nu})^{\dot{\alpha}}_{\dot{\beta}} \bar{\psi}_{\nu}^{\dot{\beta}}\right)\nonumber\ee
\be + \frac{i}{2} \chi^{\alpha}\partial_{\mu}\sigma^{\mu}_{\alpha \dot{\alpha}}\bar{\chi}^{\dot{\alpha}}+\frac{i}{2} \bar{\chi}_{\dot{\alpha}}\partial_{\mu}\bar{\sigma}^{\dot{\alpha}\alpha}\chi_{\alpha} -\frac{F^{\dagger}}{\sqrt{3}\bar{M}_{Pl}}\chi^{\alpha}\chi_{\alpha}-\frac{F}{\sqrt{3}\bar{M}_{Pl}}\bar{\chi}_{\dot{\alpha}}  \bar{\chi}^{\dot{\alpha}}\nonumber
\ee
\be 
+ \frac{1}{2\sqrt{6}\bar{M}_{Pl}}i\sigma^\mu_{\alpha\dot{\alpha}}\bar{\chi}^{\dot{\alpha}}(S_\mu^{\alpha})_{\text{matter}}+\frac{1}{2\sqrt{6}\bar{M}_{Pl}} i\chi^{\alpha} \sigma^{\mu}_{\alpha \dot{\alpha}}(\bar{S}^{\dot{\alpha}}_{\mu})_{\text{matter}}\nonumber 
\ee
\be
+\frac{iF}{\sqrt{2}\bar{M}_{Pl}}\psi^{\alpha}_{\mu}\sigma^{\mu}_{\alpha \dot{\alpha}}\bar{\chi}^{\dot{\alpha}}+\frac{iF^{\dagger}}{\sqrt{2}\bar{M}_{Pl}} \bar{\psi}_{\mu \dot{\alpha}}\bar{\sigma}^{\mu \dot{\alpha}\alpha}\chi_{\alpha}
\nonumber
\ee
\be
+\frac{1}{2\bar{M}_{Pl}}(\psi_{\mu}^\alpha (S^{\mu}_{\alpha})_\text{matter}+\bar{\psi}_{\mu\alphad}(S^{\mu\alphad})_\text{matter})\nonumber
\ee
\be
+\frac{\chi^{\alpha}}{\sqrt{2}F}\partial_{\mu}S^\mu_{\alpha \text{matter}}+\frac{\bar{\chi}_{\dot{\alpha}}}{\sqrt{2}F^{\dagger}}\partial_{\mu} \bar{S}^{\mu \dot{\alpha}}_{\text{matter}}.
\ee
\newline
The super-Higgs mechanism is realised by applying the shift
\bea
\Psi_{\mu\alpha}&\rightarrow&\psi_{\mu\alpha}-\frac{i}{\sqrt{6}}\sigma_{\mu\alpha\dot{\alpha}}\bar{\chi}^{\dot{\alpha}}-\sqrt{\frac{2}{3}}\frac{1}{m_{\frac{3}{2}}}\partial_\mu\chi_\alpha\\
\bar{\Psi}_{\mu}^{\alphad}&\rightarrow&\bar{\psi}_{\mu}^{\alphad}-\frac{i}{\sqrt{6}}\bar{\sigma}_{\mu}^{\alpha\dot{\alpha}}\chi_{\alpha}-\sqrt{\frac{2}{3}}\frac{1}{m^{\dagger}_{\frac{3}{2}}}\partial_\mu\bar{\chi}^{\alphad}
\eea
so that the Gravitino eats the Goldstino degrees of freedom and the Lagrangian becomes that of the massive Gravitino coupled to matter
\be
\mathcal{L}= -\frac{1}{2}\epsilon^{\mu\nu \rho \sigma}(\Psi_{\mu}^{\alpha}\sigma_{\nu\alpha\dot{\alpha}}\partial_{\rho}\bar{\Psi}_{\sigma}^{\dot{\alpha}}  -\bar{\Psi}_{\mu\dot{\alpha}}\bar{\sigma}_{\nu}^{\dot{\alpha}\alpha}\partial_{\rho}\Psi_{\sigma \alpha})
-i\left(m_{3/2}\Psi_\mu^\alpha(\sigma^{\mu\nu})_\alpha^\beta\Psi_{\nu\beta}+m_{3/2}^{\dagger}\bar{\Psi}_{\mu\dot{\alpha}}(\bar{\sigma}^{\mu\nu})^{\dot{\alpha}}_{\dot{\beta}}\bar{\Psi}_{\nu}^{\dot{\beta}}\right)\nonumber
\ee
\be+\frac{1}{2\bar{M}_{Pl}}(\Psi_{\mu}^\alpha (S^{\mu}_{\alpha})_{\text{matter}}+\bar{\Psi}_{\mu\alphad}(S^{\mu\alphad})_{\text{matter}}),
\ee
with the Gravitino now carrying the right degrees of freedom for the self-interaction term to be correctly identified with the Gravitino mass.  

This review of the super-Higgs mechanism generates the same Lagrangian as that of Deser-Zumino \cite{Deser:1977uq}, which is a combination of the A-V action \cite{Volkov:1973ix} plus linear supergravity, with one addtion:  In their paper they additionally comment on the two cosmological constants $-\frac{f^2}{2}e + ce$, which are set to cancel.  where $e\equiv det(e^a_{\mu})$ and $e^{a}_{\mu}$ is the Vielbein of the Graviton. In this review and in \cite{Rychkov:2007uq} this is implicitly assumed.

\providecommand{\href}[2]{#2}\begingroup\raggedright\endgroup


\begin{thebibliography}{10}

\bibitem{Salam:1974zb}
A.~Salam and J.~A. Strathdee, {\it {On Goldstone Fermions}},  {\em Phys. Lett.}
  {\bf B49} (1974) 465--467.

\bibitem{Volkov:1973ix}
D.~V. Volkov and V.~P. Akulov, {\it {Is the Neutrino a Goldstone Particle?}},
  {\em Phys. Lett.} {\bf B46} (1973) 109--110.

\bibitem{Deser:1977uq}
S.~Deser and B.~Zumino, {\it {Broken Supersymmetry and Supergravity}},  {\em
  Phys. Rev. Lett.} {\bf 38} (1977) 1433.

\bibitem{Komargodski:2009rz}
Z.~Komargodski and N.~Seiberg, {\it {From Linear SUSY to Constrained
  Superfields}},  {\em JHEP} {\bf 0909} (2009) 066,
  [\href{http://xxx.lanl.gov/abs/0907.2441}{{\tt arXiv:0907.2441}}].

\bibitem{Kuzenko:2010ni}
S.~M. Kuzenko, {\it {Variant supercurrents and Noether procedure}},  {\em
  Eur.Phys.J.} {\bf C71} (2011) 1513,
  [\href{http://xxx.lanl.gov/abs/1008.1877}{{\tt arXiv:1008.1877}}].

\bibitem{Kuzenko:2010ef}
S.~M. Kuzenko and S.~J. Tyler, {\it {Relating the Komargodski-Seiberg and
  Akulov-Volkov actions: Exact nonlinear field redefinition}},  {\em
  Phys.Lett.} {\bf B698} (2011) 319--322,
  [\href{http://xxx.lanl.gov/abs/1009.3298}{{\tt arXiv:1009.3298}}].

\bibitem{Kuzenko:2011tj}
S.~M. Kuzenko and S.~J. Tyler, {\it {On the Goldstino actions and their
  symmetries}},  {\em JHEP} {\bf 1105} (2011) 055,
  [\href{http://xxx.lanl.gov/abs/1102.3043}{{\tt arXiv:1102.3043}}].

\bibitem{Dudas:2011kt}
E.~Dudas, G.~von Gersdorff, D.~Ghilencea, S.~Lavignac, and J.~Parmentier, {\it
  {On non-universal Goldstino couplings to matter}},  {\em Nucl.Phys.} {\bf
  B855} (2012) 570--591, [\href{http://xxx.lanl.gov/abs/1106.5792}{{\tt
  arXiv:1106.5792}}].

\bibitem{Antoniadis:2011xi}
I.~Antoniadis, E.~Dudas, and D.~Ghilencea, {\it {Goldstino and sgoldstino in
  microscopic models and the constrained superfields formalism}},  {\em
  Nucl.Phys.} {\bf B857} (2012) 65--84,
  [\href{http://xxx.lanl.gov/abs/1110.5939}{{\tt arXiv:1110.5939}}].

\bibitem{Rychkov:2007uq}
V.~S. Rychkov and A.~Strumia, {\it {Thermal production of gravitinos}},  {\em
  Phys. Rev.} {\bf D75} (2007) 075011,
  [\href{http://xxx.lanl.gov/abs/hep-ph/0701104}{{\tt hep-ph/0701104}}].

\bibitem{Komargodski:2010rb}
Z.~Komargodski and N.~Seiberg, {\it {Comments on Supercurrent Multiplets,
  Supersymmetric Field Theories and Supergravity}},  {\em JHEP} {\bf 07} (2010)
  017, [\href{http://xxx.lanl.gov/abs/1002.2228}{{\tt arXiv:1002.2228}}].

\bibitem{Cheung:2010mc}
C.~Cheung, Y.~Nomura, and J.~Thaler, {\it {Goldstini}},  {\em JHEP} {\bf 03}
  (2010) 073, [\href{http://xxx.lanl.gov/abs/1002.1967}{{\tt
  arXiv:1002.1967}}].

\bibitem{Argurio:2011hs}
R.~Argurio, Z.~Komargodski, and A.~Mariotti, {\it {Pseudo-Goldstini in Field
  Theory}},  {\em Phys.Rev.Lett.} {\bf 107} (2011) 061601,
  [\href{http://xxx.lanl.gov/abs/1102.2386}{{\tt arXiv:1102.2386}}].

\bibitem{Argurio:2011gu}
R.~Argurio, K.~De~Causmaecker, G.~Ferretti, A.~Mariotti, K.~Mawatari, {\em
  et.~al.}, {\it {Collider signatures of goldstini in gauge mediation}},
  \href{http://xxx.lanl.gov/abs/1112.5058}{{\tt arXiv:1112.5058}}.

\bibitem{Bagger:2007gm}
J.~A. Bagger and A.~F. Falk, {\it {Decoupling and Destabilizing in
  Spontaneously Broken Supersymmetry}},  {\em Phys.Rev.} {\bf D76} (2007)
  105026, [\href{http://xxx.lanl.gov/abs/0708.3364}{{\tt arXiv:0708.3364}}].

\bibitem{Ferrara:1974pz}
S.~Ferrara and B.~Zumino, {\it {Transformation Properties of the
  Supercurrent}},  {\em Nucl. Phys.} {\bf B87} (1975) 207.

\bibitem{Dumitrescu:2011zz}
T.~T. Dumitrescu and Z.~Komargodski, {\it {Aspects of supersymmetry and its
  breaking}},  {\em Nucl.Phys.Proc.Suppl.} {\bf 216} (2011) 44--68.

\bibitem{Weinberg:2000cr}
S.~Weinberg, {\it {The quantum theory of fields. Vol. 3: Supersymmetry}}, .
  Cambridge, UK: Univ. Pr. (2000) 419 p.

\bibitem{Stelle:1978ye}
K.~Stelle and P.~C. West, {\it {Minimal Auxiliary Fields for Supergravity}},
  {\em Phys.Lett.} {\bf B74} (1978) 330.

\bibitem{Ferrara:1978em}
S.~Ferrara and P.~van Nieuwenhuizen, {\it {The Auxiliary Fields of
  Supergravity}},  {\em Phys.Lett.} {\bf B74} (1978) 333.

\bibitem{Cremmer:1978hn}
E.~Cremmer, B.~Julia, J.~Scherk, S.~Ferrara, L.~Girardello, {\em et.~al.}, {\it
  {Spontaneous Symmetry Breaking and Higgs Effect in Supergravity Without
  Cosmological Constant}},  {\em Nucl.Phys.} {\bf B147} (1979) 105.

\bibitem{Ellis:2011mz}
J.~Ellis and N.~E. Mavromatos, {\it {On the Role of Space-Time Foam in Breaking
  Supersymmetry via the Barbero-Immirzi Parameter}},  {\em Phys.Rev.} {\bf D84}
  (2011) 085016, [\href{http://xxx.lanl.gov/abs/1108.0877}{{\tt
  arXiv:1108.0877}}].

\bibitem{Ogievetsky:1980qp}
V.~Ogievetsky and E.~Sokatchev, {\it {EQUATION OF MOTION FOR THE AXIAL
  GRAVITATIONAL SUPERFIELD}},  {\em Sov. J. Nucl. Phys.} {\bf 32} (1980) 589.
  [Yad.Fiz.32:1142-1151,1980].

\bibitem{Ferrara:1978rk}
S.~Ferrara, M.~T. Grisaru, and P.~van Nieuwenhuizen, {\it {POINCARE AND
  CONFORMAL SUPERGRAVITY MODELS WITH CLOSED ALGEBRAS}},  {\em Nucl.Phys.} {\bf
  B138} (1978) 430.

\bibitem{Witten:1981nf}
E.~Witten, {\it {Dynamical Breaking of Supersymmetry}},  {\em Nucl. Phys.} {\bf
  B188} (1981) 513.

\bibitem{Cremmer:1978iv}
E.~Cremmer, B.~Julia, J.~Scherk, P.~van Nieuwenhuizen, S.~Ferrara, {\em
  et.~al.}, {\it {SuperHiggs Effect in Supergravity with General Scalar
  Interactions}},  {\em Phys.Lett.} {\bf B79} (1978) 231.

\bibitem{Bolz:2000fu}
M.~Bolz, A.~Brandenburg, and W.~Buchmuller, {\it {Thermal production of
  gravitinos}},  {\em Nucl.Phys.} {\bf B606} (2001) 518--544,
  [\href{http://xxx.lanl.gov/abs/hep-ph/0012052}{{\tt hep-ph/0012052}}].

\end{thebibliography}

\end{document}